# Excitonic fine structure in emission of linear carbon chains

Stella Kutrovskaya[*,1,2,3], Anton Osipov[3,4], Stepan Baryshev[5],
Anton Zasedatelev[5], Vlad Samyshkin[3], Sevak Demirchyan[1,2], Olivia Pulci[6],
Davide Grassano[6], Lorenzo Gontrani[6], Richard R. Hartmann[7], Mikhail E. Portnoi[8,9],
Alexey Kucherik[3], Pavlos Lagoudakis[5], and Alexey Kavokin[1,2,10]

[1]*School of Science, Westlake University, Hangzhou 310024, China*
[2]*Institute of Natural Sciences, Westlake Institute for Advanced Study, Hangzhou 310024, China*
[3]*Department of Physics and Applied Mathematics, Stoletov Vladimir State University, Vladimir 600000, Russia*
[4]*ILIT RAS — Branch of FSRC "Crystallography and Photonics" RAS, Shatura 140700, Russia*
[5]*Skolkovo Institute of Science and Technology, Moscow 121205, Russia*
[6]*Department of Physics, University of Rome Tor Vergata, I-00133 Rome, Italy*
[7]*Physics Department, De La Salle University, 0922 Manila, Philippines*
[8]*Physics and Astronomy, University of Exeter, Exeter EX4 4QL, United Kingdom*
[9]*ITMO University, St. Petersburg 197101, Russia*
[10]*Spin Optics Laboratory, St. Petersburg State University, St. Petersburg 198504, Russia*

E-mail: stella.kutrovskaya@westlake.edu.cn

**Abstract**

We studied monoatomic linear carbon chains stabilised by gold nanoparticles attached to their ends and deposited on a solid substrate. We observe spectral features of straight chains containing from 8 to 24 atoms. Low temperature PL spectra reveal characteristic triplet fine-structures that repeat themselves for carbon chains of different lengths. The triplet is invariably composed of a sharp intense peak accompanied by two broader satellites situated 15 and 40 meV below the main peak. We interpret these resonances as an edge-state neutral exciton, positively and negatively charged trions, respectively. The time-resolved PL shows that the radiative lifetime of the observed quasiparticles is about 1 ns, and it increases with the increase of the length of the chain. At high temperatures a non-radiative exciton decay channel appears due to the thermal hopping of carriers between parallel carbon chains. Excitons in carbon chains possess large oscillator strengths and extremely low inhomogeneous broadenings.

*Keywords:* carbon chains, excitons, photoluminescence spectra, nanoparticles Several types of low-dimensional crystals based on carbon were attracting attention of the physical and chemical research communities in the XXI century. Nanodiamonds, fullerenes, carbon nanotubes and graphene demonstrate a variety of interesting and unusual electronic properties that make them promising for a variety of applications in nano-electronics and photonics.[1] One of the most challenging goals for the nanofabrication technology is the realization of ultimate one-dimensional crystals, monoatomic chains of sp-carbon. Traces of two stable allotropes of sp-carbon (polyyne and cumulene) have been found in nature: in meteorite craters, interstellar dust, natural graphite and diamond mines.[2-5] The high chemical reactivity of linear acetylenic carbon and its low stability at room temperature and atmospheric pressure makes it difficult to extract free standing carbon chains from natural sources. Moreover, multiple attempts to synthesize polyyne chains artificially yielded modest or no success so far.[6] Until now, to the best of our knowledge, stable freestanding samples of straight carbon chains have not yet been realised.

challenge as, in general, infinite one-dimensional atomic chains are unstable in vacuum. According to the Landau theorem,[7] fluctuations prevent the formation of ideal one-dimensional crystals. Therefore, many works have been devoted to the artificial stabilization of carbon chains. The stabilization may be achieved by the use of heavy anchor atomic groups.[8] The fabrication of carbon chains *in — situ* TEM allowed producing the chains of about 5 nm long.[9] The synthesis of carbon chains inside double-wall nanotubes[10] was shown to be an efficient way for the realization of macroscopically long carbon isolated from the environment. In order to fabricate free-standing chains the pinning to metal surfaces has been used. [11] The stabilization of carbon chains by tris(3,5-di-t-butylphenyl)methyl moiety led to the observation of a record 44 atom long free-standing chain.[12] One-dimensional carbon crystals are expected to exhibit unique mechanical, optical and electronic properties.[13] According to the recent theoretical works,[14] one-dimensional carbon chains could form the most robust of all known crystals. It may take one of two allotropic forms: cumulene, where neighboring carbon atoms are linked with double electronic bonds, and polyyne, where single and triple electronic bonds alternate (see the schematic in see Fig. 1a.). Note that here we mean by polyyne a series of consecutive alkynes, (—C = C—) $_n$ with n greater than 1 that is not necessarily ended by hydrogen atoms. In polyyne, the interatomic distances are predicted to be 0.133nm (C — C) and 0.123 nm (C = C) which yields the lattice constant of 0.256 nm. These interatomic separations are significantly less than those in graphite where the spacing between neighbouring atoms is of 0.142 nm, while the spacing between atomic planes is 0.335 nm. This is why the predicted Young modulus of linear carbon chains is somewhat higher than in graphene and one order of magnitude higher than that of diamond.[15] The electrical conductivity and magnetic properties of carbon chains can be dramatically varied by stretching and twisting them.[16] The strain induced switching of conductivity in one-dimensional carbon crystals has been predicted[17] and observed recently. [18] Isolated monoatomic chains are predicted to demonstrate peculiar spin-dependent quantum transport effects.[19] Sp-carbon crystals are also expected to exhibit superconductivity at high temperatures.[20,21] Ab-initio calculations[14] show that infinite-chain polyyne is a direct band gap semiconductor, with a gap of 2.58 eV. Our own ab-initio calculations of the band structure of stabilised infinite polyyne chains predict the existence of a direct electronic band-gap of 2.7 eV at the edge of the Brillouin zone see Fig. 1b. Being direct-gap semiconductors, polyyne chains are expected to possess unusual optical properties. In particular, their non-linear optical response is expected to be giant.[22] Indeed, unlike graphite or graphene which are strong absorbers and do not emit light, and carbon nanotubes where the multi-valley band structure leads to dark excitons suppressing the luminescence, one-dimensional carbon crystals are attractive because of their ability to emit visible light. Room temperature photoluminescence (PL) spectra of carbon chains reported in a work[23] demonstrate a sequence of broad maxima corresponding to the chains of different lengths. The band gap of finite size linear chains dramatically depends on the number of carbon atoms in a chain, being the larger the shorter the chain. In the following we shall focus on straight polyyne chains containing from 8 to 24 carbon atoms whose band structure is reduced to a sequence of discrete energy levels similar to molecular orbitals. The gap between the highest occupied molecular orbital (HOMO) and the lowest unoccupied molecular orbital (LUMO) varies in the range of 2-4 eV (see the schematic in see Fig. 1d). This results in a large spectral distribution of allowed optical transitions in such systems.

In this Letter, we study stable polyyne chains synthesised by the laser ablation (LAL) technique in a colloidal solution. The mechanical stabilisation of sp-carbon is achieved due to the electron bonding of carbon chains to gold nanoparticles (NPs). When deposited on a substrate, the stabilized chains demonstrate straight parts whose lengths significantly exceed the theoretical limit for a free stable monoatomic carbon chain. The high-resolution transmission electron microscopy (HR TEM) of our samples shows straight linear carbon chains of the lengths that sometimes exceed 5 nm. Most interestingly, at the liquid Helium temperature, a pronounced fine structure emerges in the photoluminescence (PL) spectra of the deposited carbon chains on the top of broad resonances corresponding to polyyne chains of different lengths. This fine structure is composed invariably by a high amplitude narrow peak accompanied by two lower energy satellites which can be attributed to the exciton resonance and two trion resonances, respectively. The time-resolved photoluminescence (TRPL) spectra show that the radiative lifetime of the observed transitions is of the order of 1 ns, that is similar to the data reported for excitons in CNTs.[24] At high temperatures, the double exponential decay of the PL signal is observed with a fast exponential characterising the thermal dissociation of excitons, and the slow exponential corresponding to their radiative decay. The exciton radiative lifetime decreases with the decrease of the length of the chain. We refer to the Su-Schrieffer-Heeger model[25] to argue that the transition that dominates low temperature PL spectra is based on the edge electronic states that form the HOMO-LUMO pair in carbon chains stabilised by gold NPs. This interpretation is confirmed by the ab-initio calculation (Fig. 5e,f,g).

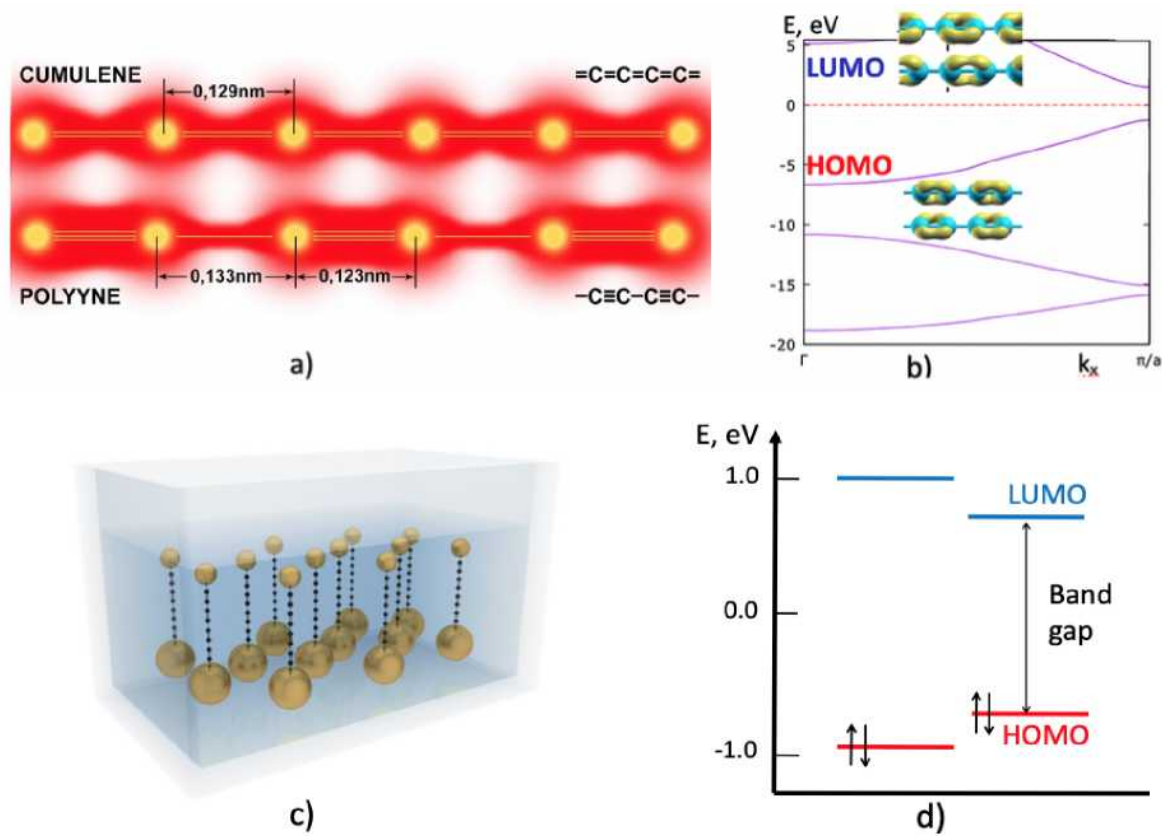

Figure 1: The structure of monoatomic carbon chains: (a) the schematic distribution of the electron density in cumulene and polyyne allotropes. (b) shows the band structure of an infinite polyyne chain predicted by our ab-initio calculation, and the HOMO, LUMO doubly degenerate orbitals. (c) illustrates the concept of stabilisation of monoatomic carbon chains by gold nanoparticles (NPs) in a solution. Gold NPs are shown out of the actual scale. (d) schematically shows the energy level structure of a finite polyyne chain composed by 14 carbon atoms with two gold nanoparticles attached to its ends.

**Results. The synthesis of monoatomic carbon chains.** We employed the LAL method for synthesis of the linear carbon chains.[34] The laser processing resulted in the formation of polyyne threads.[29] The stabilization of linear carbon chains is achieved by adding spherical gold NPs to the solution.[27] The procedure is schematically illustrated in Fig. 1c. NPs having average sizes of either 10 or 100 nm attach themselves to the ends of carbon chains by single electron bonds. As also folding of chains and formation of kinks occurs predominantly at the single bonds, the parity rule is imposed to the number of atoms belonging to straight parts of the chains. Here, in particular, we observe spectral resonances of the straight chains containing even numbers of C atoms ranging from 8 to 24. Typically, these are straight parts of longer linear chains attached by both ends to gold NPs. Kinks separate each linear chain into several straight parts. It is important to note also that if NPs at the opposite ends of a carbon chain are of different sizes, the difference of their work functions results in the charging of the carbon-NP complex that acquires a stationary dipole moment. This dipole polarization provides a tool for the chain ordering by an applied voltage.[28]

We intentionally ordered the carbon chains by passing the solution through the stationary electric field in the course of sputtering (see the scheme in Fig. 2a[28]). As a result, we were able to deposit long parallel chains of polyyne. Fig. 2c shows the HR TEM image of a typical bundle of carbon chains attached by its both ends to gold NPs. Fig. 2d shows the HR TEM image of the central part of the bundle of parallel carbon chains of the length exceeding 40 nm. Gold NPs remained outside the frame of the image in this case. Our previous study showed that the ensemble of carbon chains in a bundle forms a kind of one-dimensional Van- der-Waals crystal, where the distance between neighbouring chains exceeds the interatomic distance in a single chain d by a factor of 3.6,[28] approximately (see Fig. 2b). We note that only about one half of carbon bundles have gold NPs of different sizes (10 nm and 100 nm, typically) at their ends. These bundles are dipole polarised due to the difference of the work function of gold NPs of different sizes. They are aligned by the electric field during the sputtering process. These structures are hosts to positively and negatively charged trions that manifest themselves in the low-temperature PL spectra as we show below. The other half of carbon-metal nanostructures are formed by gold NPs of the same size. They are not polarized and not aligned by the electric field. The carbon chains in these structures mostly host electrically neutral excitons. The carbon chains connecting gold NPs are of approximately the same length, which is why they are able to form hexagonal quasicrystal structures. This is confirmed by the X-ray study presented in Ref.[28]. The kinks, indeed, have different locations at different carbon chains. In Figure 2 (c,d) we show HR-TEM images of various carbon bundles typically composed by over 10 parallel chains.

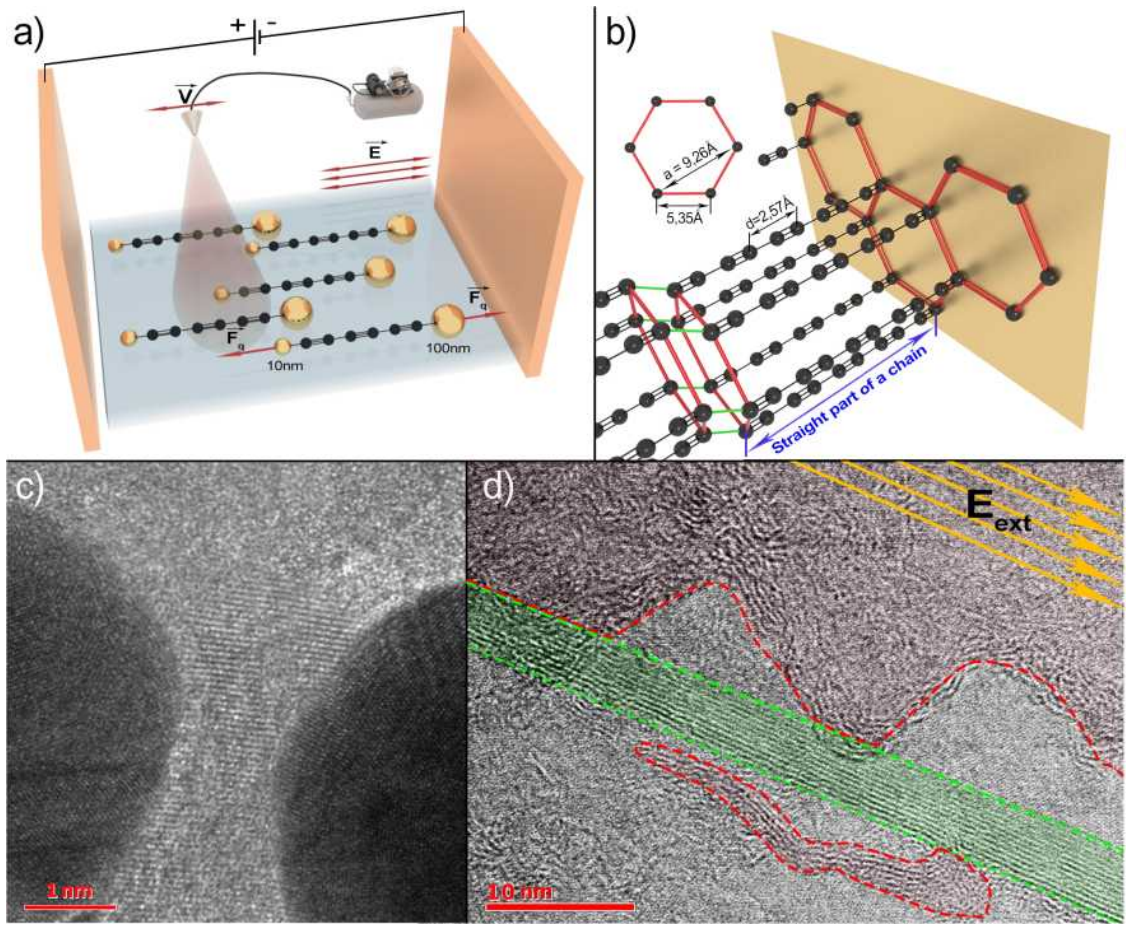

Figure 2: Deposition and characterisation of carbon chains: (a) illustrates our method of deposition of the aligned carbon chains stabilised by gold NPs of different sizes. Due to the difference in work functions of gold nanoparticles attached to the ends of a carbon chain, it becomes dipole polarised. The external electric field orients the polarised chains, so that over a half of them appears to be aligned when deposited on a substrate by sputtering, (b) shows the schematics of the structure of a quasicrystal formed by polyyne wires attached to a gold NP (shown as a yellow wall) that is based on the X-ray diffraction analysis;[28] (c) shows the HR TEM image of bundles of carbon chains attached to two gold NPs of a nearly spherical shape, seen as the dark regions on the left-hand side and right-hand sides of the image. The observed quasi-crystal structure is composed by sp-carbon polyyne chains and represents a Van-der-Waals quasicrystal. Its lattice parameters are a =0.926 nm, c = 1.25 nm and the polyyne lattice constant is d =0.2558 nm. In this case the distance between neighboring chains in the hexagon is 0.535 nm, in full agreement with the HR TEM image of the central part of the field oriented bundle shown in (d). The green area corresponds to the field-aligned polarised chains endcaped with Au NPs, the red area is occupied by randomly oriented unpolarized chains.

**Low-temperature PL spectra of the deposited polyyne chains.** In the PL experiments, we excited the deposited polyyne chains quasi-resonantly, at the wavelengths between 370 and 390 nm. The strong emission at the wavelengths over 400 nm has been detected. As there were necessarily thousands of individual carbon chains under our pump spot, which is of 2 ^m size, the PL signal is always built from contributions of carbon chains of different lengths. The chains containing from 8 to 18 atoms[34] typically dominate the spectra. Fig. 3a shows the characteristic PL spectra featuring broad resonances corresponding to the chains of different lengths emitting in the spectral range from 2.4 to 3.0 eV. The lowest energy optical transition shifts to the red with the increase of the length of the chain, in full agreement with the theoretical predictions.[34] In Fig. 3a we attribute the sequence of spectral resonances observed to the polyyne chains containing from 8 to 22 atoms (the transitions in chains of 8 to 18 atoms are labeled in the figure) in agreement with the previously reported transition energies[34] and the results of our own ab-initio calculation. As the temperature goes down to 4K, a very distinct and peculiar fine structure emerges on the top of each broad PL peak (Fig. 3b). Nearly identical triplet structures are observed for the chains of 10, 12, 14, 16 atoms. A strong and very narrow peak with a full width at half maximum (FWHM) of only 3-4 meV has two broader and lower satellites shifted by about 15 and 25 meV towards lower energies. This observation is in a stark contrast with the exciton spectra of carbon nanotubes[24] that typically feature strongly inhomogeneously broadened peaks with typical FWHM of several tens meV. The left panel of Fig. 4 shows the time- resolved photoluminescence (TRPL) data acquired at the high temperature (Fig. 4a) and liquid Helium temperature (Fig. 4c). The TRPL is spectrally resolved, the selected bands being indicated in the insets. The high temperature spectra show the double-exponential behaviour reflecting the interplay between non-radiative (presumably, thermal hopping) and radiative channels of the exciton decay. The extracted decay times are plotted in Fig. 4b. The low temperature spectra show the monoexponential decay of the excitonic photoluminescence with a characteristic decay time of the order of 1 ns that is comparable with the typical radiative lifetimes of excitons in carbon nanotubes.[35,36] The exciton radiative lifetime increases with the increase of the length of the chain following the anticipated behaviour of the matrix element of the dipole transition. We note that the exciton fine structure observed in CW PL spectra taken at the selective optical excitation cannot be resolved in the TRPL spectra as the pulsed excitation used in this spectroscopy technique is not wave-length selective. Still, the method allows for clearly distinguishing between the non-radiative exciton decay due to the thermal hopping of carriers and the exciton radiative recombination that dominates the low-temperature TRPL dynamics.

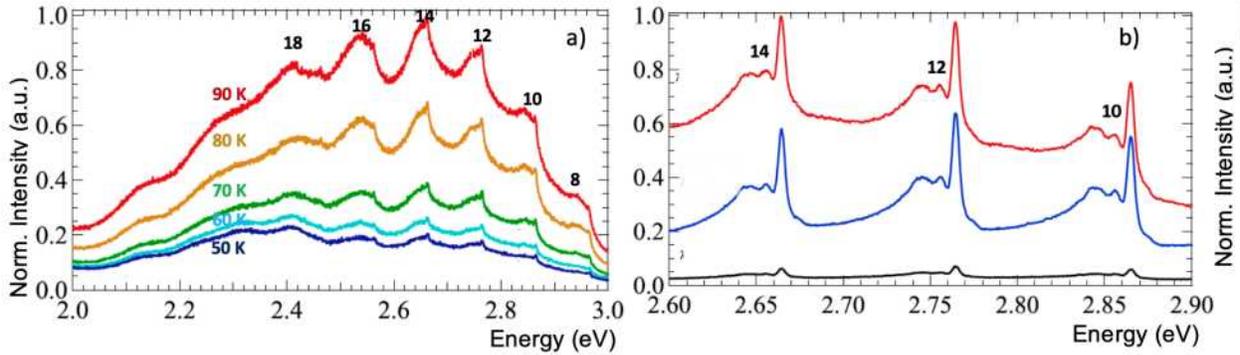

Figure 3: PL spectra of the deposited polyyne chains of different lengths (the number of atoms in the chain is indicated on the top of the corresponding spectral resonance): (a) shows the spectra taken at temperatures from 90 to 50K (red curve corresponds to 90 K, yellow curve corresponds to 80K, green curve corresponds to 70K, teal curve corresponds to 60K, blue curve corresponds to 50K). The laser excitation wavelength is 390 nm with the intensity of 5 mW and the acquisition time of 10 s. (b) shows the PL spectra taken at 4K. Red, blue and black curves correspond to the excitation wavelengths of 390, 380 and 370 nm, respectively. The acquisition time is 40 s.

**Discussion.** The triplet structures emerging as the temperature decreases are indicative of excitonic transitions. It is important to note that the textbook definition of the exciton binding energy as the difference of the transition energy for an electron-hole pair separated by an infinite distance and the exciton optical transition energy cannot be applied in our case, as electron and hole cannot be separated by more than a few nanometers in a Van-der-Waals nanocrystal composed of a handful of carbon chains. The quantity that is relevant to our experimental observations is rather an exciton thermal dissociation energy. It characterises the difference of the transition energies between an electron and a hole belonging to the same carbon chain and between an electron and a hole located in different (neighboring) carbon chains. The exciton dissociation into a radiatively inactive electron-hole pair by thermal hopping of one of the carriers to the neighboring chains (Fig. 2c) is expected to govern the temperature dependence of the intensity of the observed excitonic peaks. From the PL spectra taken at temperatures between 4 and 300 K, we estimate the thermal hopping energy to be of the order of 15-20 meV. This is consistent with our variational calculation performed using the procedure developed in[37] for quantum wires. Using the electron and hole effective masses predicted by our ab-initio calculation as $m_e = 0.078 m_0$, $m_h = 0.09 m_0$ with $m_0$ being a free electron mass, describing both dielectric and metallic screening by an effective dielectric constant eps = 6 for spatially direct and *eps* = 4 for spatially indirect excitons, we obtain the difference of energies of an exciton formed by an electron and a hole belonging to the same chain and an exciton formed by an electron and a hole belonging to different chains about 15 meV. Given the strongly limited accuracy of the variational calculation, we find these numbers to be in a very good agreement with the experimental data. Note that the excitons are expected to be localised at the ends of the carbon chains forming an equivalent of Su-Schrieffer-Heeger states[25] as confirmed by the results of our ab-initio calculation shown in see Fig. 5e, f, g.

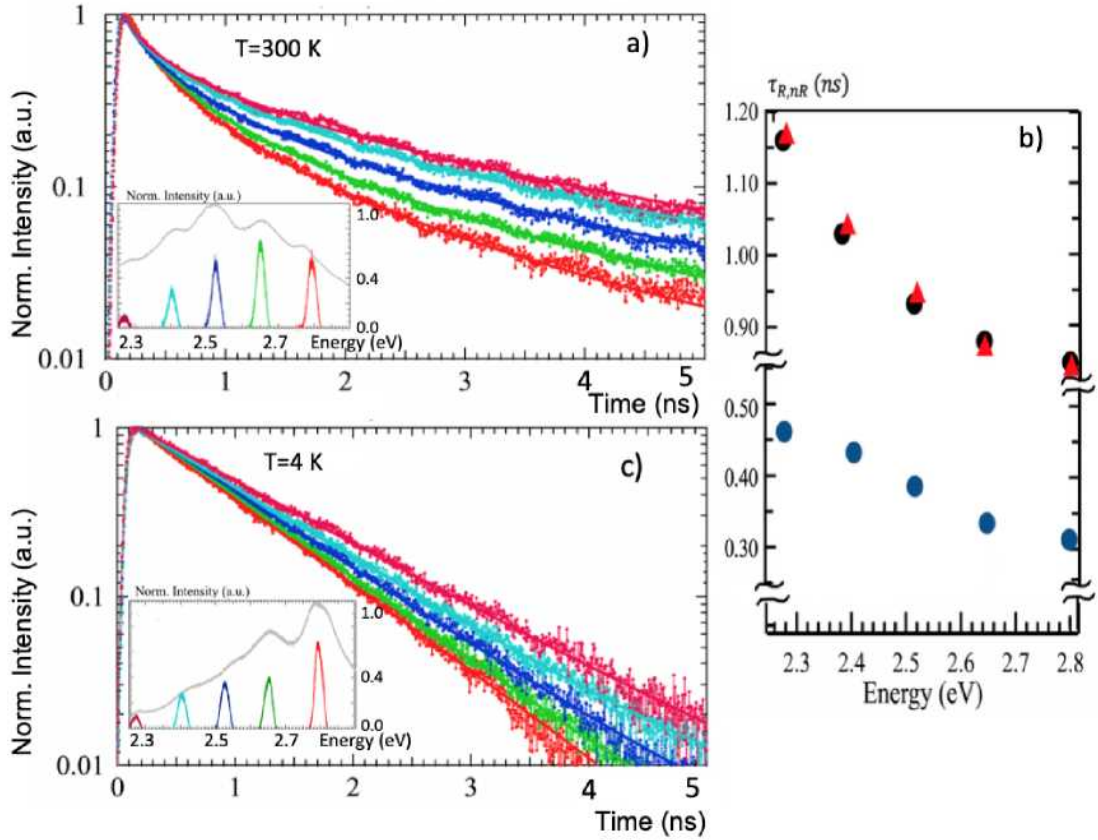

Figure 4: The time-resolved photoluminescence (TRPL) data: (a) and (c) show the TRPL signal acquired at the room temperature and at 4K, respectively. The insets show the spectral bands that correspond to the TRPL curves of (a,c), respectively. The colors match.(b) shows the extracted decay times of the TRPL signal taken at room temperature and Helium temperature, respectively. Red and black points correspond to the deduced radiative decay time at the room temperature (red) and Helium temperature (black). The blue points show the non-radiative decay times, extracted from the room temperature TRPL curves (a).

Two lower energy satellites clearly visible in the excitonic spectra of the carbon chains containing 10, 12 and 14 atoms may be associated to zero-phonon lines,[38] phonon replicas, bi-excitons or charged excitons, in principle. We rule out the three first scenario as (i) the measured vibron energies in polyyne are of 130 and 267 meV (Raman spectra are shown in the Supplementary Material, figure S3) that strongly exceeds the observed splittings between the main peak and satellite peaks (15-25 meV), (ii) the typical optical phonon energy in polyyne is expected to be of the order of 100 meV[36] that is also significantly larger than the observed splittings, and (iii) the bi-exciton line intensity would vanish at low excitation powers, that contradicts our data. Charged exciton complexes (trions) are likely to be formed in our system due to the vicinity of metal nanoparticles that can supply carbon chains with extra charge carriers, see the schemes in Fig. 5b,c and TEM images given in the Supplementary Material. The variational calculation predicts the exciton-positively charged trion splitting of about 15 meV and positive-negative trion splitting of 25 meV which is in a good agreement with the data. The strong dipole polarisation of approximately a half of gold-stabilised carbon chains is confirmed by their alignment in the external electric field (see the scheme in Fig. 3a and Supplementary Material), thus the presence of extra-charges localised at the ends of some of the chains is beyond any doubts. The spectrally integrated intensities of negatively charged (X-) and positively charged (X+) trion transitions are expected to be lower than the integrated intensity of the neutral exciton transition. Indeed, statistically, having equal concentrations of 10 nm and 100 nm NPs, one should expect only a half of carbon chains to be attached to the NPs of different radii and hence polarised. In the polarised chains two of the lowest energy optical transitions would be either a positively or a negatively charged trion. As every chain is divided into several straight pieces by kinks, neutral excitons can still be formed in central parts of the polarised chains. We conclude that the sum of integrated intensities of X- and X+ lines should be equal or less that the integrated intensity of the X-line. This is generally consistent with the data (see Fig. 5d).

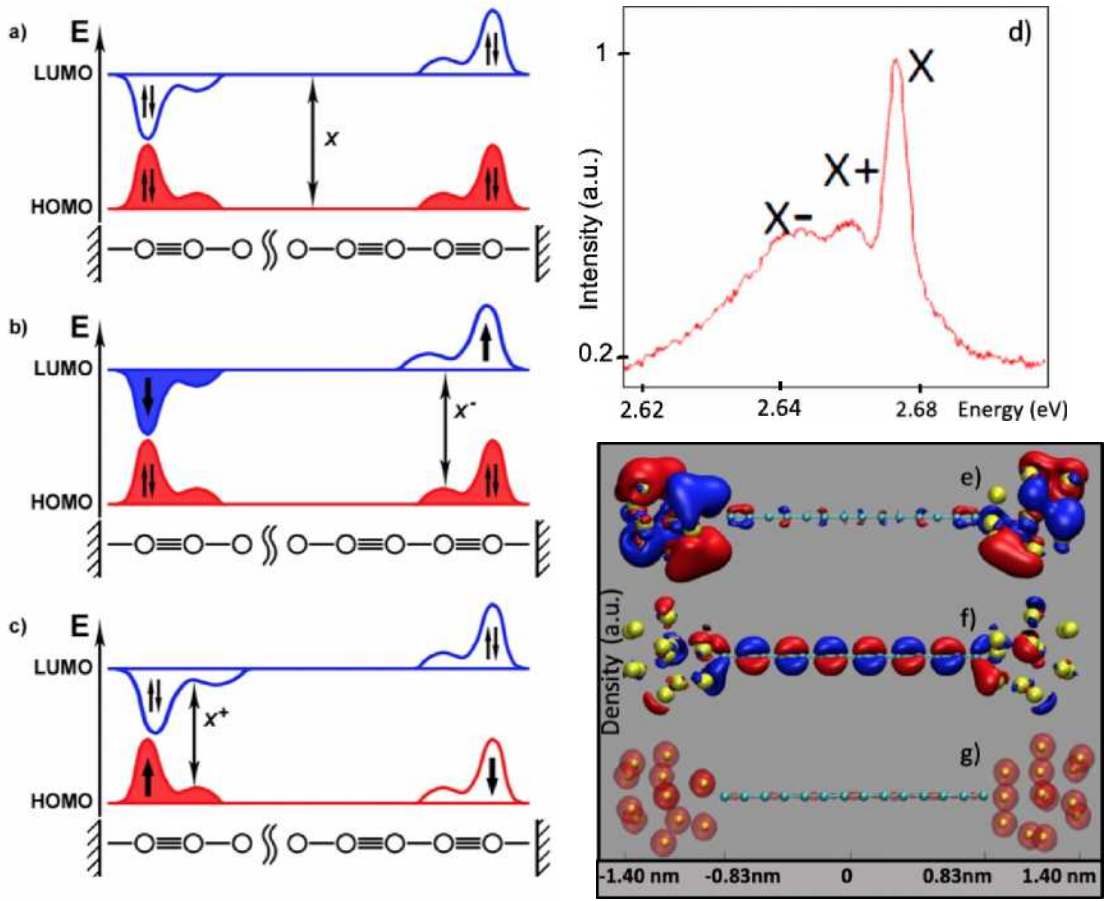

Figure 5: The scheme of excitonic transitions in a finite-size polyyne chain. (a) The neutral exciton (X) is formed by even and odd edge states that originate from the HOMO-LUMO pair. (b,c) The trion transitions in charged chains, where the left-right symmetry is broken so that the optical transition occurs between either the original HOMO state and the spin polarised electron state localised at one of the edges (X-) or between the hole state localised at one of the edges and the original LUMO state (X+). (d) The enlarged PL spectrum of a characteristic triplet corresponding to the 14-atom chain that shows the relative strengths of exciton and trion transitions. Molecular orbitals and total electron density plots for a model system containing a polyyne chain composed of 14 carbon atoms capped with two gold NPs ($C_{14}$-$Au_{13_2}$) for HOMO (e), LUMO (f). The total calculated electron density (g). Red and blue lobes correspond to positive and negative values, respectively.

Fig. 5e,f,g presents the plots of HOMO and LUMO wavefunctions and of the total electron density of Au-stabilized carbon chains, obtained from *in vacuo* DFT calculations with localised basis sets, are shown for the polyyne chain of 14 carbon atoms capped with two Au13 nanoparticles. Indeed, in the calculated distributions of the HOMO and the total density (top and bottom panel) it appears that the most part of the electron cloud is located around gold atoms and partially on the shorter C-C bonds of the polyyne moiety. Excitons and trions based on HOMO-LUMO transitions are localised at the ends of the atomic chains. This strong spatial localisation and significant electron-hole overlap are responsible for a large oscillator strength of the corresponding optical transitions. The similar plots obtained for C14 without gold terminations (see Figure S6 and S7 of the Supplemental Material) lack these features, and the electronic density tends to vanish moving from the centre to the ends of the chain. We conclude that gold-stabilized carbon chains are highly advantageous for the observation of excitonic effects as they provide a very strong spatial localisation, large oscillator strength and low inhomogeneous broadening of the exciton and trion transitions.

In conclusion, we show that the sharp peaks emerging at low temperatures in the PL spectra of gold-stabilised carbon chains are indicative of the exciton and trion transitions based on the edge electronic states in the chains. The triplet fine structure that is very well seen at Helium temperature is essentially independent of the length of the chain, while the absolute energies of the transitions become larger for the shorter chains. This observation demonstrates a high potentiality of synthesised polyyne chains for opto-electronic applications, especially in nano-lasing and single photon emitters. Moreover, the observation of radiatively active excitons in ultimate one-dimensional carbon crystals is of a great fundamental interest. Further studies are needed to fully reveal properties of excitons and trions in carbon chains.

**Methods.** We have used the method of laser fragmentation of colloidal carbon systems for realisation of stabilised carbon chains. To create carbon-gold bonds we additionally illuminated the solution by nanosecond laser pulses generated by an Ytterbium (Yb) fiber laser having the central wave length of 1.06 mm, the pulse duration of 100 ns, the repetition rate of 20 kHz and the pulse energy of up to 1 J. The time between subsequent pulses was about 1 s. The sizes of gold NPs have been controlled by the dynamic laser scattering device Horiba SZ100. The formation of polyyne carbon phase we have controlled on the Raman spectra measured by using the Senterra spectrometer, made by Bruker company. The pump laser wavelength is of 532 nm at the power level of 40 mW, the radiation was focused through a 50-fold microlens, the spectra were collected in the confocal microscope configuration and averaged over 10 measurements. The accumulation time of each measurement was 60 seconds. For the detailed study of the orientation distribution of nanodipoles, we have performed the high resolution transmission electron microscopy and X-ray diffraction studies using FEI Titan$^3$ with a spatial resolution of up to 2 A . Processing of TEM-images and diffraction patterns was conducted with the opened database package Image J 1.52 a. We study the photoluminescence (PL) of carbon chains using the 140 fs pulsed 80 MHz repetition rate Ti: Sapphire laser system (Coherent Cameleon Ultra ll) coupled to an optical parametric oscillator allowing for fine control over the excitation energies. As the sample is mounted in a closed-cycle cryostation, Montana Instruments Cryostation C2, we carry out PL measurements in a vacuum within the temperature range of 4K-300K. The optical excitation we employ in these experiments is tightly focused onto a sample by Mitutoyo M Plan APO SL x50 microscope objective with a numerical aperture of 0.42. Note that we use three different excitation wavelengths to elucidate the nature of the observed fine lines in the PL spectra, namely 390, 380 and 370 nm as shown in Figure 3(b). We collect PL signal in the transmission configuration using the same Mitutoyo M Plan APO SL x50 microscope objective. To avoid any parasitic signal in PL spectra we filter out optical excitation with a short pass filter cutting light above 561nm, and in the collection path we use a long pass filter transmitting light above 400nm to get rid of the optical excitation signal. Being properly filtered, the PL from the sample is coupled into a 750 mm focal length spectrometer (Princeton Instruments SP2750) equipped with a charge-coupled device camera (Pixis-XB: 1024BR) with 1024x1024 imaging array. In all the spectroscopic measurements we use 1200 grooves mm$^{-1}$ grating and 20 ^m entrance slit, as well as step and glue technique to record the entire PL spectra represented in Figures 3 and 4. The spectral resolution of our setup is 100 pm. Despite omnipresent background emission from the substrate (above 650 nm) we systematically observe several distinct peaks in the PL of the sample that correspond to the optical transitions of carbon chains of the different lengths. We further extend our spectroscopic analysis with in-depth study of the PL dynamics attributed to the identified optical transitions. We employ a standard time-correlated single-photon counting technique (TCSPC), with detection wavelengths scanned across the full emission bandwidth using a variable birefringent filter device (Liquid crystal tunable filter Varispec). The recorded spectra of the PL coupled to TCSPC system are shown in Fig. 4a and Fig. 4b. We extract spectral lines within 5 nm linewidth that corresponds to the exciton and trion transitions. We use a single-photon avalanche photodiode (APD) (id100-MMF50) together with TCSPC module (SPC-160, Becker & Hickel GMBH) to detect photon events related to the PL. The typical count rate is $10^3$-$10^4$ counts/s while a dark count rate is 40 Hz. The APD response defines the time resolution in our system, which is as short as 60 ps. By using the aforementioned technique, we measure emission decay for five individual optical transitions (Fig. 4(a,b,c,d)).

The Density Functional Theory calculations for the infinite chain were performed using the QUANTUM ESPRESSO package.[39] We tested several exchange and correlation potentials: LDA, LSDA, PBE, PBE0, HSE06. An energy cutoff of 90 Ry and a k-points mesh of 16x1x1 k-points was used to optimize the atomic positions. A supercell with a vacuum of about 30 A was employed, in order to avoid spurious interactions between neighbouring chains. We found that the polyyne geometry (that is, an infinite chain with single and triple bonds) can only be obtained with the use of hybrid pseudopotentials. In LDA, LSDA or PBE, instead, cumulene is the only stable structure, with a zero electronic gap. Optical spectra for varying lengths of the carbon chains (from 10 up to 26 C atoms) were obtained within Time Dependent Functional Theory (TDDFT) calculations with PBE0 kernel, using the code Gaussian[40] with triple-zeta 6-311+G(d,p) basis set plus LANL2DZ pseudopotential for gold core electrons,[41] after geometry optimisation. In all the considered cases, polyyne configurations have a lower energy than cumulene.

# Acknowledgement

This work of S.K. and A.KA. is from the Innovative Team of International Center for Polari- tonics and is supported by Westlake University (Project No. 041020100118). We acknowledge the support from the Program 2018R01002 funded by Leading Innovative and Entrepreneur Team Introduction Program of Zhejiang. This work was also partially supported by RFBR grants 17-32-50171, 18-32-20006, 19-32-90085, 20-52-12026, 20-02-00919 and the RScF grant 20-72-10145. A.KA. acknowledges Saint-Petersburg State University for the research grant ID 40847559. The financial support from the EU MSCA RISE projects CoExAN (GA 644076) and 'DiSeTCom' (GA823728) is acknowledged. The work of A.O. was partially supported by Ministry of Science and Higher Education within the State assignment FSRC Crystallography and Photonics RAS (Project No 075-00842-20-00). M.E.P. and R.R.H. acknowledge support from the EU MSCA RISE project TERASSE (GA 823878). The work of M.E.P. was supported by the Government of the Russian Federation through the ITMO Fellowship and Professorship Program 5-100. L.G. acknowledges support from Regione Lazio, through Progetto di Ricerca 85-2017-15125 according to L. R. 13/08. CPU computing time was granted by CINECA HPC center and by the "Departments of Excellence-2018" Program (Dipartimenti di Eccellenza) of the Italian Ministry of Research, DIBAF-Department of University of Tuscia, Project 'Landscape 4.0 - food, wellbeing and environment". The synthesis and deposition of LLCC have been performed at the Vladimir State University. Raman spectra and absorbance were measured at the Center for Optical and Laser Materials Research, Research Park, St. Petersburg State University. TEM measurements have been performed in the 'System for microscopy and analysis" LLC, Moscow.

## Supporting Information Available

The detailed description of the synthesis of linear carbon chains in a liquid, experimental method of measurements, the orientated deposition of linear carbon chains-gold nanostructures; the high resolution TEM image of a reference sample where the deposition was performed in the absence of external electric field and the ab-initio calculation results for a non-stabilised carbon chain.

## Author Contributions

S.K. has conceived the work and analysed experimental data; A.O. contributed to the laser ablation experiments; S.B., A.Z. and P.L. contributed to the measurements and analysis of the PL and time-resolved PL spectra; V.S. realized the LLCC deposition on a substrate; S.D. contributed to the variational calculation of exciton states; O.P. contributed to the ab-initio model and revised the manuscript; D.G. performed the ab-initio calculation for an infinite chain; L.G. performed the ab-initio calculation for finite chains; R.H. contributed to the fit of the PL spectra; M.E.P. contributed to the theoretical model and interpretation of the results; A.K. contributed to the synthesis of monoatomic carbon chains and realized HR TEM microscopy study; A.KA. contributed to the interpretation of the results and has coordinated the collaborative work; S.K. and A.KA. have written the manuscript.